\documentclass[11pt,twocolumn]{article}
\pdfoutput=1
\usepackage{cite}
\usepackage{amssymb,amsmath,mathrsfs}
\usepackage{graphicx}
\usepackage[latin1]{inputenc}
\usepackage[dvipsnames]{xcolor}
\usepackage[footnotesize]{caption}
\usepackage{xcolor}
\usepackage{hyperref}
\usepackage[hmarginratio=1:1,top=32mm,columnsep=20pt]{geometry}
\usepackage{abstract}

\def\gsim{\mathrel{\rlap{\lower4pt\hbox{\hskip1pt$\sim$}}
    \raise1pt\hbox{$>$}}}

\title{\vspace{-15mm}%
	\fontsize{16pt}{10pt}\selectfont
	\textbf{Clues on the Majorana scale from scalar resonances at the LHC}
	}	
\author{%
	\large
	Oliver~Fischer\footnote{E-mail: \texttt{oliver.fischer@unibas.ch }}\\[10pt]
	\normalsize	Department of Physics, University of Basel, \\ 
	\normalsize 	Klingelbergstr.\ 82, CH-4056 Basel, Switzerland\\[5pt]
	\vspace{-5mm}
	}
\date{}

\begin{document}

\twocolumn[\begin{@twocolumnfalse}
\maketitle

\vspace{-10pt}
\hrulefill
\vspace{-5pt}
\begin{abstract}
\noindent 
In order to address the observation of the neutrino oscillations and the metastability of the Standard Model, we extend the fermion sector with two right-handed (i.e.\ sterile) neutrinos, and the scalar sector of the SM with a real scalar, the Hill field.
The latter takes the r\^{o}le of a Majoron and generates the Majorana masses for the neutrino sector, such that the particle spectrum features two CP-even scalars $h_1$ and $h_2$, and also two heavy, mass degenerate neutrinos.
When the $h_1$ is identified with the scalar resonance at $\sim$125 GeV and the condition is imposed that the $h_1$ self coupling and its running vanish at the Planck scale, the scalar mixing and the vacuum expectation value of the Hill field are fixed by the $h_2$ mass.

The $h_2$ can be searched for at the LHC, and it has prospects of being discovered for the target integrated luminosities of the HL-LHC and the Future Circular hadron Collider (FCC-hh) when its mass is on the weak scale.
The knowledge of the $h_2$ mass and its decay properties can yield an insight into its coupling to the heavy neutrinos, and thus also on the heavy neutrino mass scale. 
This yields an interesting connection between potentially detectable heavy scalars in high-energy proton collisions and the mass scale of the heavy neutrinos that is testable at the LHC and at future colliders.
\end{abstract}
\vspace{-10pt}
\hrulefill
\vspace{10pt}
\end{@twocolumnfalse}]

\saythanks

\section{Introduction}
\noindent
Despite its remarkable success in describing data at different energy scales, the Standard Model (SM) of particle physics has some built-in shortcomings, two of which we are going to address in the following:

Firstly, it does not provide a renormalisable way to generate the light neutrinos' masses, which requires an extension, for instance, with right-handed neutrinos. In this class of models a so-called Majorana mass term, involving exclusively the sterile neutrinos, and a Yukawa coupling term, connecting sterile neutrinos with the three active neutrinos and the Higgs doublet, are possible. The scale of the Majorana mass is generally not predicted by the theory, and thus free to reside anywhere between zero and the Planck scale.

The second built-in shortcoming of the SM is the stability of the electroweak vacuum. The present central values of the Higgs boson mass, $m_h = 125.09 \pm (0.21)_{\rm stat} \pm (0.11)_{\rm syst}$ GeV \cite{Aad:2013wqa,Chatrchyan:2013mxa,Aad:2015zhl}  and the top quark mass, $m_t = 173.34 \pm (0.27)_{\rm stat} \pm (0.71)_{\rm syst}$ GeV \cite{ATLAS:2014wva} suggest that the running of the quartic Higgs self-coupling becomes negative at the renormalisation scale $\Lambda \sim 10^{11}$ GeV when the two-loop renormalisation group equations for the Standard Model \cite{Degrassi:2012ry,Branchina:2013jra,Bezrukov:2012sa} are used,  which renders the Higgs potential metastable.

As was shown in refs.\ \cite{Gonderinger:2009jp,Basso:2013nza} singlet extensions of the scalar sector can help to control the evolution of the Higgs self coupling, such that the vacuum can be stabilised. Furthermore it is possible to have the self coupling vanish exactly at the Planck scale, which removes the need of new energy scales between the Fermi and Planck scales, and allows for the possibility that the electroweak scale is determined by Planck physics \cite{Bezrukov:2012sa}. Moreover, the question of vacuum stability can be connected to the dark matter in extensions with scalar singlets and scalar $SU(2)_L$ doublets \cite{Khan:2014kba,Khan:2015ipa}.

Aside from its shortcomings, the present agreement between the SM theory prediction and precision measurements results in strong, direct and indirect constraints on all New Physics models. In particular the neutrino Yukawa couplings are strongly constrained via the non-unitarity of the PMNS matrix to be at most ${\cal O}(10^{-2})$ for heavy neutrino masses ${\cal O}(100)$ GeV \cite{Antusch:2014woa}. Also additional scalar singlets are strongly constrained by the measurement of the Higgs boson properties and precision data, see e.g.\ refs.\ \cite{Bechtle:2008jh,Bechtle:2015pma} and references therein.

While right-handed neutrinos are difficult to detect in direct searches at the LHC \cite{Deppisch:2015qwa}, scalar fields show better prospects to be detected and are being thoroughly searched for \cite{Robens:2015gla,Robens:2016xkb,vonBuddenbrock:2016rmr,Dupuis:2016fda}. No clear signal of a heavy scalar particle has yet been found, unless one counts the recent diphoton excess \cite{diphoton} and the statistical signal with three sigma significance reported in ref.\ \cite{vonBuddenbrock:2015ema}.

In the following, we introduce a minimal framework that can generate the light neutrinos' masses, and can address the metastability of the SM. The model is defined by extending the scalar sector of the SM with one real scalar singlet $H$, and $n=2$ right-handed singlet fermions $N$ (cf.\ refs.\ \cite{Gelmini:1980re,Schechter:1981cv}). The condition, that the SM-like Higgs boson self coupling and its running vanish at the Planck scale, fixes the scalar mixing as a function of the mass of a heavy scalar boson. This also fixes the scale of the Majorana mass, and makes the model very predictive.

\section{The scalar sector}
It was shown in ref.\ \cite{Schechter:1981cv} that the combination of a scalar Higgs doublet and additional scalar fields, with a non-zero lepton number and couplings to the neutral fermions, can account successfully for the neutrino masses.
More recently the effect of such a complex scalar field on the stability of the electroweak vacuum has been investigated in ref.\ \cite{Bonilla:2015kna} in the context of the neutrino mass mechanism.

Here we consider the scalar sector of the SM to be extended with one real scalar singlet, which, for simplicity, we parametrize with the Hill Higgs model, as introduced in ref.\ \cite{Hill:1987ea} and studied e.g.\ in refs.\ \cite{Basso:2012nh,Zhang:2015uuo}:
\begin{equation}
V = - \frac{\lambda _1}{8} \left( \Phi^\dagger \Phi - v^2\right) ^2 - \frac{\lambda _2}{8} \left( \sqrt{2} f_2 H - \Phi^\dagger \Phi \right) ^2 \,,
\label{def:scalarpotential}
\end{equation}
with $\Phi$ the Higgs doublet field and $H$ the scalar singlet Hill field.  A defining feature of the Hill model is the absence of $H$ self-interactions, as well as the quartic $H^2\Phi^2$ terms, which reduces the number of new parameters to two. 

In unitary gauge, it is $\Phi=\left(0, (h + v_{\rm EW})/\sqrt{2}\right)^T$, with $v_{\rm EW}=246.22$ GeV being the vacuum expectation value of the SM-like Higgs boson, and $H= (h' + v_H)/\sqrt{2}$. The two {\it CP}-even scalars $h$ and $h'$ mix as follows:
\begin{equation}
\left( \begin{array}{c} h_1\\ h_2 \end{array} \right) = \left( 
\begin{array}{cc} c_{\alpha} & s_{\alpha}\\ - s_{\alpha} & c_{\alpha} \end{array} \right) \, \left( \begin{array}{c} h \\ h^\prime \end{array} \right)\,,
\end{equation}
with $s_\alpha$ and $c_\alpha$ being the sine and cosine, respectively, of the scalar mixing angle $\alpha$.
The two mass eigenstates $h_{1}$ and $h_2$ couple to the SM particles like the SM Higgs boson with an overall prefactor of $s_{\alpha}$ and $c_{\alpha}$, respectively. The parameters $f_2$, $\lambda_2$ and $\lambda_3 := \lambda_1 + \lambda_2$ can be expressed in terms of the observable scalar masses and the mixing angle:
\begin{subequations}
\label{eq:hillparameters}
\begin{eqnarray}
\lambda_2 &=&  \frac{s^2_\alpha c^2_\alpha (m^2_{h_2} - m^2_{h_1})^2}{2\, v_\mathrm{EW}^2 (c_\alpha^2 m^2_{h_2}+s^2_\alpha m^2_{h_1})}\, ,\\
\lambda_3 &=& \lambda_{\rm SM} + s_\alpha^2 \frac{m_{h_2}^2-m_{h_1}^2}{2\,v_{\rm EW}^2}\,, \label{eq:lambda3}\\
f_2 &=& v_\mathrm{EW} \frac{c_\alpha^2 m^2_{h_1} + s^2_\alpha m^2_{h_2} }{|s_\alpha| c_\alpha (m^2_{h_2} - m^2_{h_1})}\,,
\end{eqnarray}
\end{subequations}
with $\lambda_{\rm SM}$ being the self coupling of the SM Higgs boson. Furthermore, the vacuum expectation value of the Hill field is given by 
\begin{equation}
v_H = \frac{|s_\alpha| c_\alpha (m^2_{h_2} - m^2_{h_1})}{ c_\alpha^2 m^2_{h_1} + s^2_\alpha\,m^2_{h_2}} \frac{v_{\rm EW}}{2}\,.
\label{eq:vH}
\end{equation}
\begin{figure}
\includegraphics[width=0.45\textwidth]{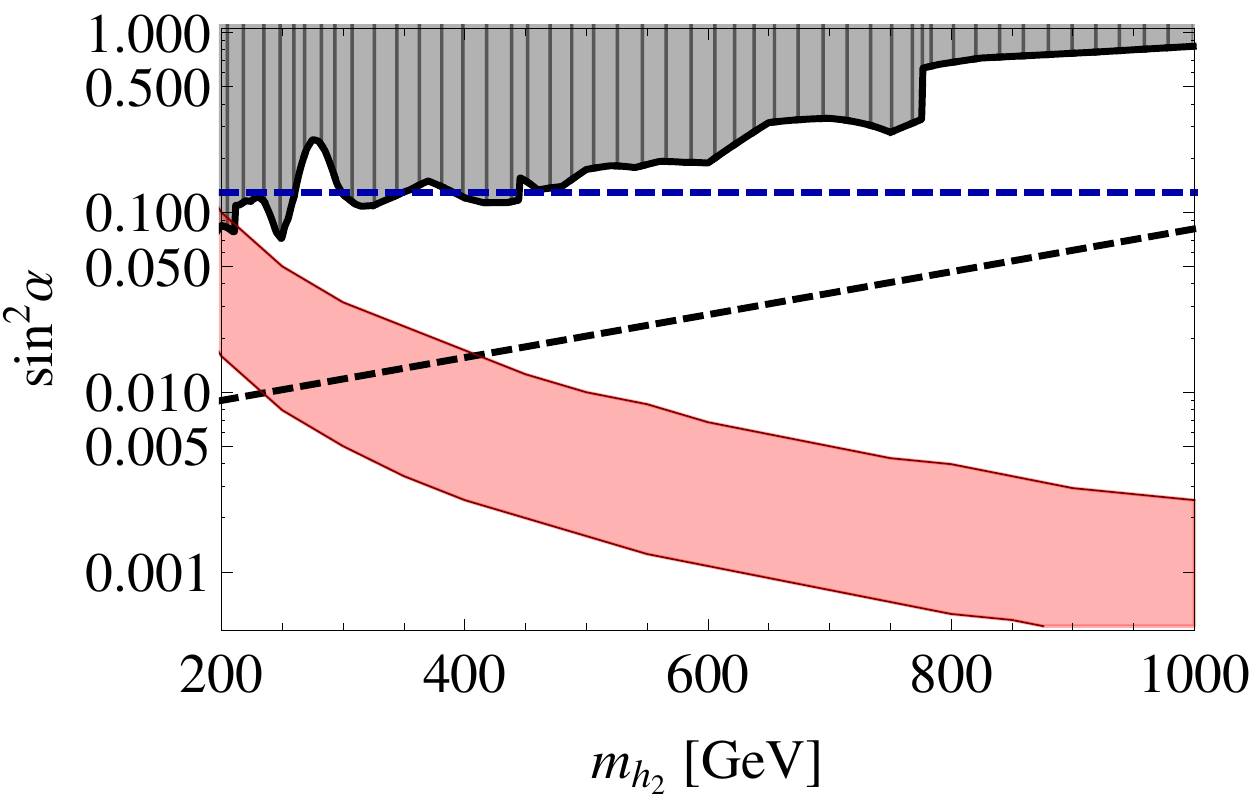}
\begin{center}
\vspace{-10pt}
\includegraphics[width=0.4\textwidth]{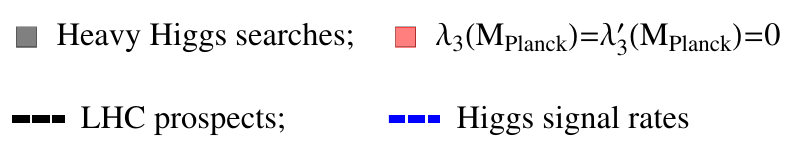}
\end{center}
\caption{The parameter space in the Hill model that results in the SM-like Higgs self coupling to vanish at the Planck scale (see \eqref{eq:condition}) is denoted by the red area. Two-loop renormalisation group equations for the SM parameters have been used and Higgs mass, top mass, and the strong coupling constant have been varied within their 2$\sigma$ bounds.
The black solid and dashed line denotes the exclusion limits from LHC direct searches \cite{Robens:2016xkb} and our extrapolated search prospects, respectively, see text for details.}
\label{fig:rge}
\end{figure}

A few remarks are in order at this point. 
Firstly, we consider the Hill model as an efficient parametrisation of an underlying, more general scalar potential. In scenarii where the scalar mixing is small, this is sufficient to capture the collider phenomenology of the additional heavy scalar field.
Secondly, we posit that no terms with odd powers in the Hill field $H$ be present, which can be achieved for instance with a suitable (global) symmetry, which is broken by the Hill field.  The term linear in $H$ can thus be renormalized to zero without spoiling the renormalizability of the model, and the term proportional to $f_2 H$ remains constant under rescalings, cf.\ ref.\ \cite{Hill:1987ea}.
Lastly, we notice that the choice of a real scalar field does not give rise to a massless Goldstone Boson after symmetry breaking.

In the following, we reproduce the results from ref.\ \cite{Basso:2013nza}, wherein the evolution of the self coupling $\lambda_{3}$ was studied, when the SM effective potential (at two loop) was extended with the one-loop contributions from the Hill model. 
It turns out, that the relevant contribution from the Hill field is the shift of the Higgs self coupling, $\delta_{\lambda}:=\lambda_3-\lambda_{\rm SM}$ (cf. eq.\ \eqref{eq:lambda3}), and that other modifications of the SM parameters from the Hill field are negligible in comparison.

For what follows we posit the condition
\begin{equation}
\lambda_3(M_{\rm Planck})=0\,, \qquad \lambda_3^\prime(M_{\rm Planck})=0\,,
\label{eq:condition}
\end{equation}
such that no additional scale for new physics is required \cite{Bezrukov:2012sa}, which allows the assumption that $m_{h_2}\leq 1$ TeV.
We show the combination of scalar mixing $s_\alpha^2$ and $m_{h_2}$ that fulfil the condition in \eqref{eq:condition} by the red area in fig.\ \ref{fig:rge}. 
The width of the red area stems from the 2$\sigma$ variation of the Higgs mass with $m_{\rm Higgs} = 125.09 \pm 0.48$ GeV, the top pole mass with $m_t = 173.34 \pm 1.52$ GeV and the strong coupling constant with $\alpha_s = 0.1185 \pm 0.0012$ \cite{Agashe:2014kda}.

The constraints on $s_\alpha^2$ from direct searches for heavy scalar singlets in LHC data \cite{Robens:2015gla,Robens:2016xkb} are denoted by the gray area in fig.\ \ref{fig:rge}. 
An indirect constraint on scalar mixing comes from the measured signal rates of the SM-like Higgs boson at $\sim$125 GeV into SM particles, which is shown by the blue dashed line.
We denote the prospects of the LHC direct searches with the black dashed line, which is obtained by smoothing out and scaling the heavy Higgs searches with a factor $\sqrt{3 \text{ ab}^{-1}/20 \text{ fb}^{-1}} \simeq 10$.

We note that the for $m_{h_2} > 300$ GeV strongest constraint on scalar mixing stems from a shift on the $W$ mass \cite{Lopez-Val:2014jva}. We neglect this constraint, however, since the shift on the $W$ mass due to scalar mixing is opposed by the shift due to leptonic mixing (cf.\ ref.\ \cite{Antusch:2014woa}), such that these two effects can cancel.

\begin{figure}
\includegraphics[width=0.45\textwidth]{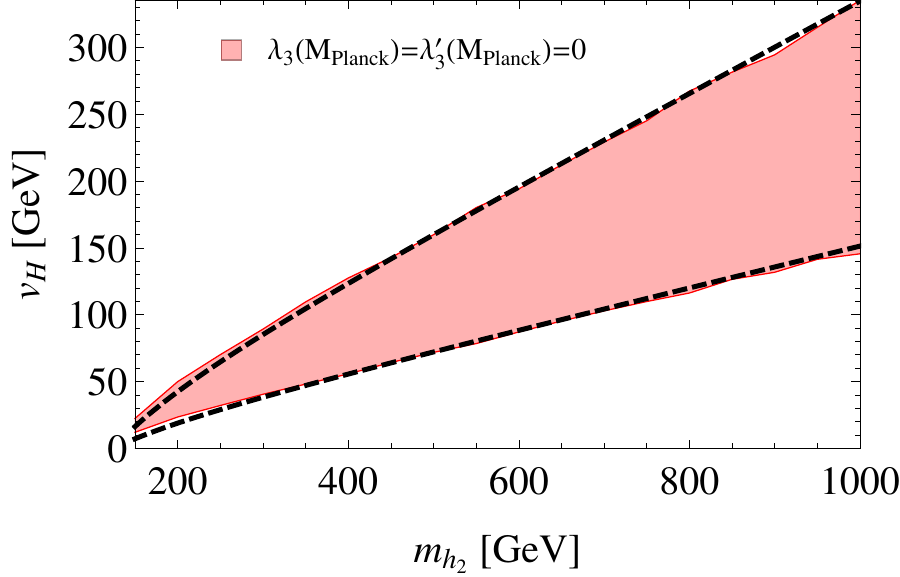}
\caption{The vacuum expectation value $v_H$ of the Hill field as a function of the $h_2$ mass, when the scalar mixing is fixed by condition \eqref{eq:condition}. The black, dashed lines correspond to the approximation in eq.\ \eqref{eq:vHsimple}.}
\label{fig:Hillvev}
\end{figure}

For $m_{h_2} \gg m_{h_1}$ the vacuum-stability can be approximated with 20 GeV $\leq s_\alpha m_{h_2} \leq 50$ GeV, whence
\begin{equation}
0.08\, \frac{v_{\rm EW}}{m_{h_1}}\, m_{h_2} \leq v_H \leq 0.17\,\frac{v_{\rm EW}}{m_{h_1}}\, m_{h_2}\,.
\label{eq:vHsimple}
\end{equation}
We show in fig.\ \ref{fig:Hillvev} the bounds on $v_H$ from the condition in eq.\ \eqref{eq:condition}, wherein the linear dependency on $m_{h_2}$ as predicted in the approximation in eq.\ \eqref{eq:vHsimple} is evident.
It is interesting to note that our results are in quantitative agreement with the stability constraints on scalar masses and mixing for the more general scalar potential in ref.\ \cite{Bonilla:2015kna}, at least in the domain where $m_{h_1}^2\ll m_{h_2}^2 \ll v_\sigma^2$, where the latter is the singlet vacuum expectation value of the general model. Differences (for neutrino Yukawa couplings equal zero) may be attributed to the variation of the top Yukawa coupling and $\alpha_s$, the different loop order for the renormalisation group running, and, of course, the different meaning of the singlet vacuum expectation value in the two models.

\section{The neutrino sector}
The seesaw mechanism with Majorana masses on the electroweak scale was introduced as ``inverse seesaw'' \cite{Wyler:1982dd,Mohapatra:1986bd}. 
Viable low-scale seesaw models with large Yukawa couplings have been introduced in ref.\ \cite{Pilaftsis:1991ug}, wherein the importance of neutrino-Higgs interactions and the stability of the light-neutrino spectrum under UV-finite radiative effects has been discussed.
Further variants of the type I seesaw mechanism that allow for heavy neutrino masses close to the electroweak scales can be found e.g.\ in refs.\ \cite{Shaposhnikov:2006nn,Kersten:2007vk,Gavela:2009cd}.

For simplicity, we consider the model in ref.\ \cite{Antusch:2015mia} for the neutrino sector in the following. We introduce $2$ right-handed neutrinos $N_1$ and $N_2$, and a ``lepton number''-like symmetry, under which the $N_1$ ($N_2)$ carries the same (opposite) charge as the SM leptons. 
This allows the following terms for the Lagrangian:
\begin{equation}
\mathscr{L} \supset y_{\nu_{\alpha}} \overline{ L}_\alpha \tilde{\Phi} N_1^{} + y_{H}\, H\, \overline{N_1^c} N_2^{}\, + \text{ H.c.}
\label{eq:lagrangian}
\end{equation}
with the flavour index $\alpha=e,\mu,\tau$, and $y_H$ real without loss of generality. Notice that in the case of $H$ carrying a global charge the term proportional to $y_H$ can still be generated if, for instance, $N_2$ is also appropriately charged.
In this way lepton number is conserved and the mass scale of the light neutrinos does not constrain the active-sterile mixing $|\theta_\alpha|$ through the observed smallness of the light neutrinos' masses.

From the Lagrangian from eq.\ \eqref{eq:lagrangian}, when $\Phi$ and $H$ develop their vacuum expectation values, the $2\times 2$ Majorana matrix $m_N$ and $3\times 2$ Dirac matrix $m_D$ emerge.
Diagonalising the resulting mass matrix of the $3+2$ neutral fermions results in the active-sterile mixing parameters:
\begin{equation}
\theta \sim m_D^{} m_N^{-1}\,,
\label{def:thetaa}
\end{equation}
if the mixing is small, i.e.\ $|\theta|=\sqrt{{\rm Tr}\left(\theta^\dagger \theta\right)}\ll 1$. Then the masses of the light and the heavy neutrinos are well approximated by:
\begin{equation}
M_{\nu_{\rm light}} \simeq -m_D^{}\, m_N^{-1} m_D^T = 0\,,\,\,  M_{\nu_{\rm heavy}} \simeq y_H\,v_H\,.
\end{equation}
We notice that the light neutrinos remain massless in the limit of the ``lepton number''-like symmetry being exact.
These masses can be introduced, when the symmetry is explicitly broken, and also when a number $n\geq 1$ of additional right-handed neutrinos $N_{(2+n)}$ are introduced. We may thus consider the additional terms
\begin{equation}
\mathscr{L}_{\rm perturpation} \supset y_{\nu_{\alpha i}}^{\prime} \overline{ L}_\alpha \tilde{\Phi} N_i^{} + y_{H_{ij}}^\prime\, H\, \overline{N_i^c} N_j^{}\, + \text{ H.c.}
\label{eq:perturbations}
\end{equation} 
with $i,j\geq 2$, $y_\nu^\prime \ll y_\nu$, and $y_H^\prime \ll y_H$, parametrising small perturbations of the protective symmetry.

In this scenario, which can be interpreted as a multi-family version of the simplistic structure in eq.\ \eqref{eq:lagrangian} (cf.\ e.g.\ refs.\ \cite{Schechter:1981cv,GonzalezGarcia:1988rw}), with the ``lepton number''-like symmetry being perturbed, the additional degrees of freedom allow for inverse seesaw (see e.g.\ ref.\ \cite{Gavela:2009cd} and references therein), and also for linear seesaw (see refs.\ \cite{Barr:2003nn,Malinsky:2005bi}).

We remark, that the contribution to the renormalisation-group evolution of $\lambda_3$ due to the mixing between active and sterile neutrinos is negligible for $y^\prime_\nu \ll y_{\nu} \leq {\cal O}(0.1)$, cf.\ refs.\ \cite{Bonilla:2015kna,Rose:2015fua}.

\section{LHC phenomenology of the $h_2$}
The phenomenology of sterile neutrinos and searches at proton colliders has been discussed e.g.\ in ref.\ \cite{Pilaftsis:1991ug} with an emphasis on the Higgs boson (see also refs.\ \cite{Kersten:2007vk,delAguila:2007qnc}).

In the following, we are considering the phenomenology of the heavy scalar, where the dominant contribution to $h_2$ production stems from gluon and vector boson fusion, respectively, due to scalar mixing. For $m_{h_2} > 200$ GeV we can approximate 
\begin{equation}
\frac{400}{m_{h_2}^2}\sigma_{\rm SM}(m_{h_2}) \leq \sigma_{h_2} \leq \frac{2500}{m_{h_2}^2}\sigma_{\rm SM}(m_{h_2})\,,
\end{equation}
with $m_{h_2}$ in GeV and $\sigma_{\rm SM}(m_{h_2})$ the production cross section of a would-be SM Higgs boson with mass $m_{h_2}$.
The right-handed neutrinos can also contribute to $h_2$ production via $Z$ and $W$ boson fusion, which is, however, suppressed with $|\theta|^4\leq {\cal O}(10^{-6})$ \cite{Antusch:2014woa}.
Moreover, the $h_2$ can be radiated off a $h_1$ in the form of $h_2$-strahlung, albeit suppressed with the virtuality of $h_1$. 

\begin{figure}
\includegraphics[width=0.45\textwidth]{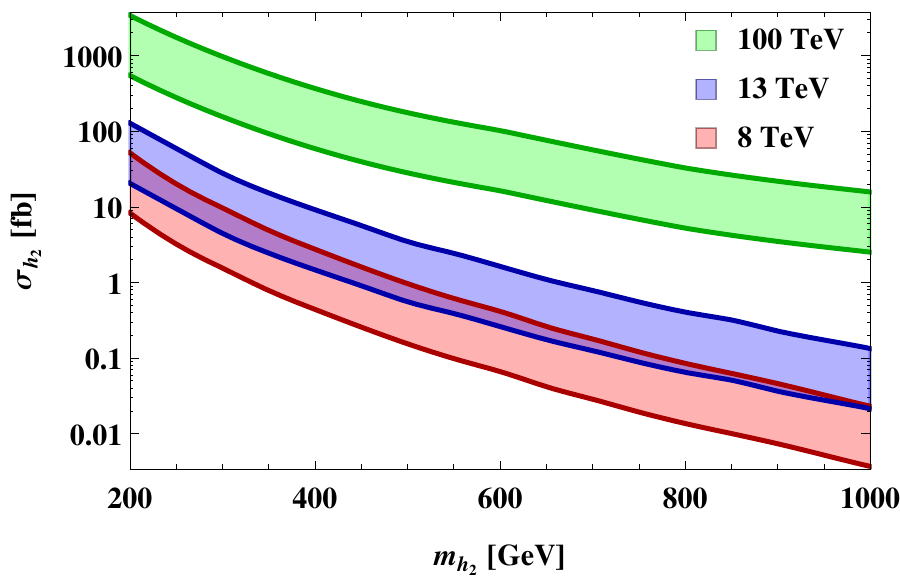}
\caption{Production cross sections for the process $pp \to h_2 j j$ for center-of-mass energies of 8, 13 and 100 TeV, as a function of the mass $m_{h_2}$. The scalar mixing for each mass is fixed by the vacuum-stability condition.  }
\label{fig:h2production}
\end{figure}

We computed the $h_2$ production cross section for the process $pp\to h_2 jj$ with WHIZARD \cite{Kilian:2007gr,Moretti:2001zz}, using the built-in parton-density functions for the proton beams $p$, and show it in fig.\ \ref{fig:h2production} as a function of $m_{h_2}$ for the center-of-mass energies of the LHC, 8 and 13 TeV, respectively. We also include the production cross section at 100 TeV for the Future Circular hadron Collider (FCC-hh) that is discussed e.g.\ in ref.\ \cite{Golling:2016gvc} (see also references therein), and which we take to be representative for the Super Proton Proton Collider \cite{Tang:2015qga}.
For the cross section, only production via scalar mixing is considered, which is fixed by the condition in \eqref{eq:condition}.
As fig.\ \ref{fig:h2production} shows, the production cross section increases by a factor of $\sim 100$ when increasing the center-of-mass energy from ${\cal O}(10)$ TeV to 100 TeV. With the planned integrated luminosity of 20 ab$^{-1}$, the FCC-hh is thus an $h_2$ factory, where several million $h_2$ can be be produced if $m_{h_2} < 1$ TeV.

\begin{figure*}
\begin{minipage}{0.4\textwidth}
\includegraphics[width=0.9\textwidth]{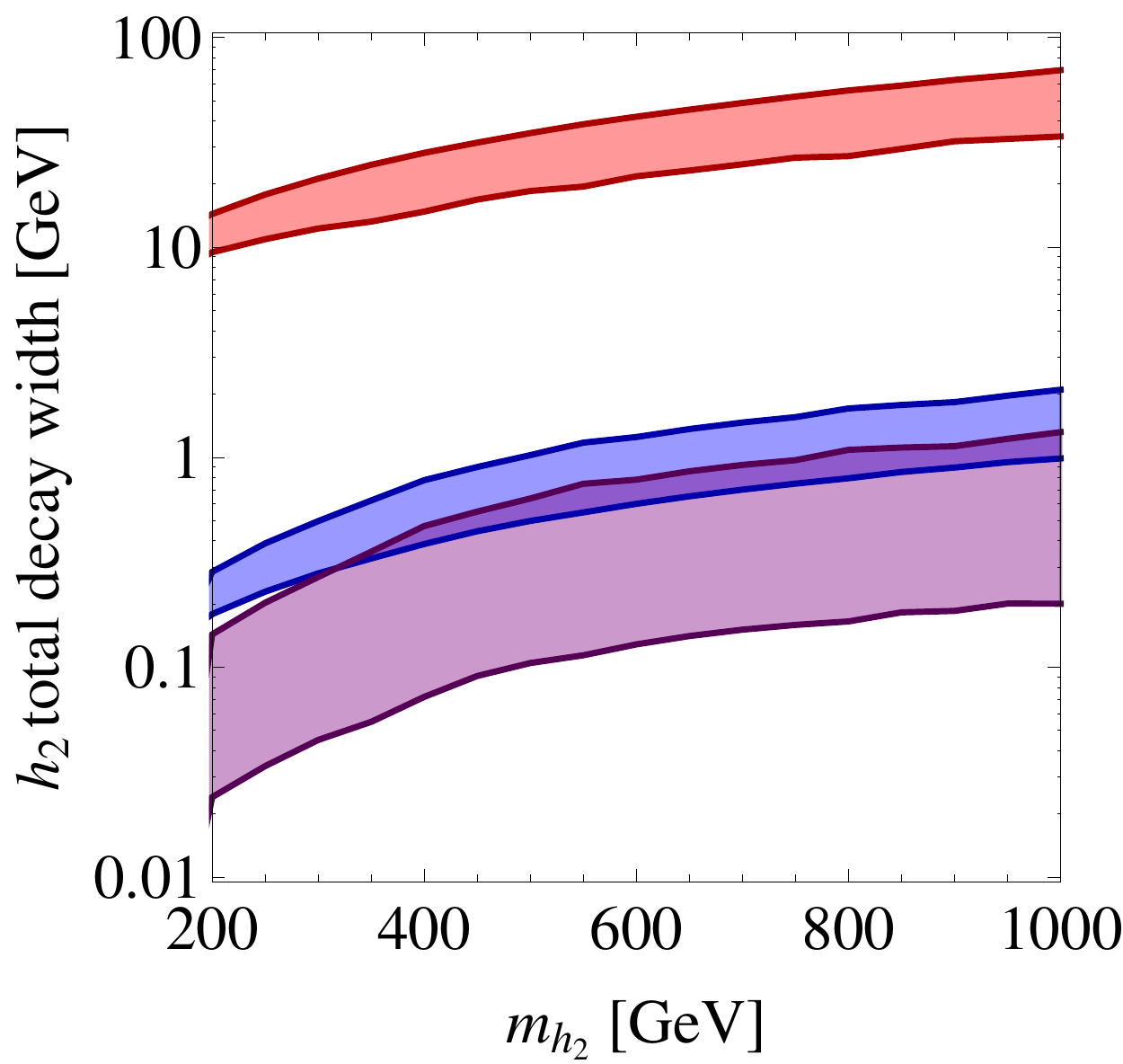}
\end{minipage}
\begin{minipage}{0.15\textwidth}
\includegraphics[width=0.9\textwidth]{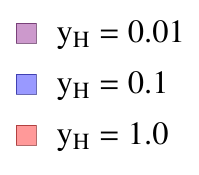}
\end{minipage}
\begin{minipage}{0.4\textwidth}
\includegraphics[width=0.9\textwidth]{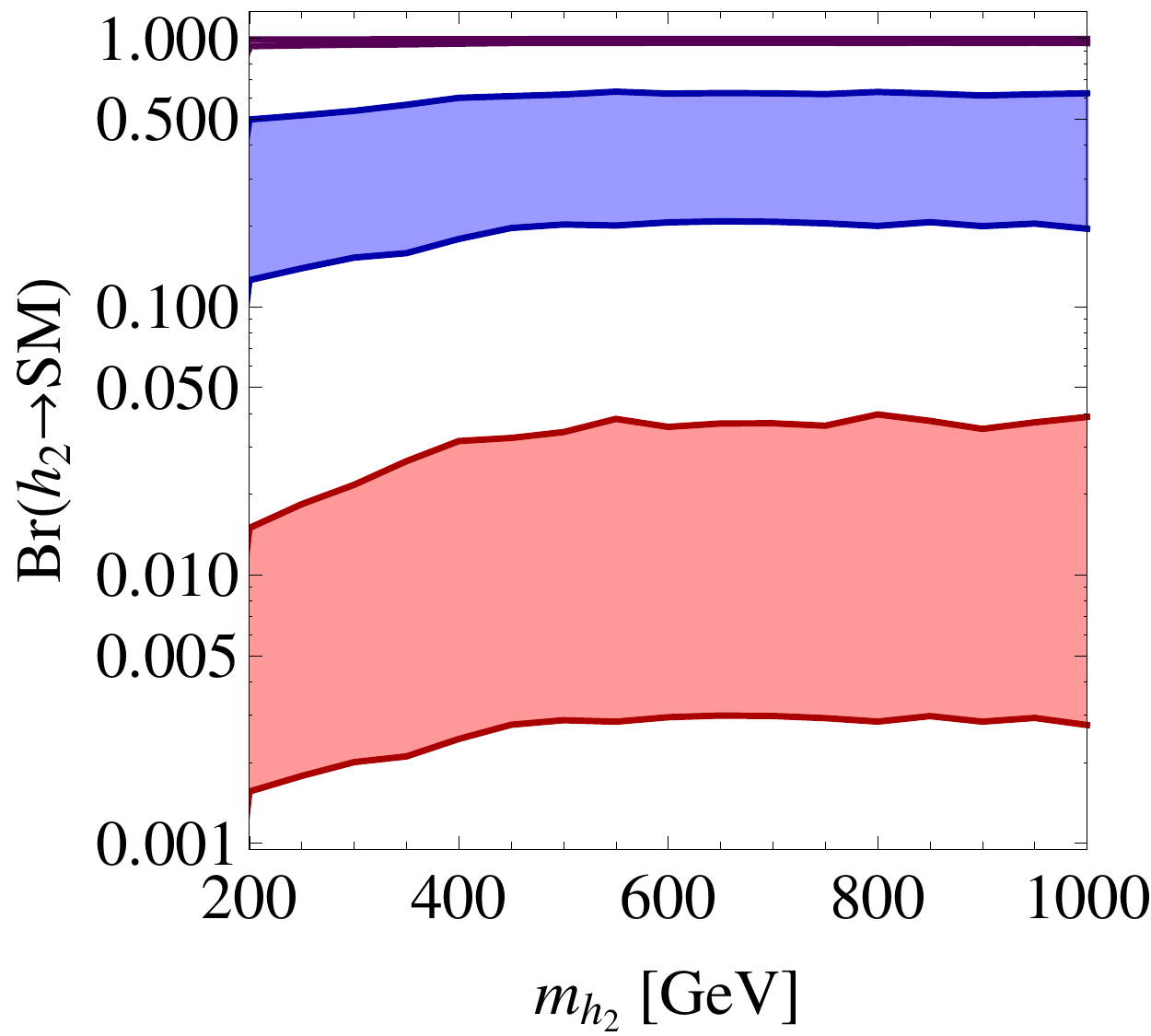}
\end{minipage}
\caption{{\it Left:} Total decay width of the heavy scalar $h_2$, as a function of its mass, with the scalar mixing fixed by the vacuum-stability condition. The colored areas denote different values for the coupling to the right-handed neutrinos $y_H$. {\it Right:} Branching ratios of the $h_2$ into SM particles, including the decay $h_2 \to h_1\,h_1$.}
\label{fig:h2-rvis}
\end{figure*}
In order to assess the detectability of the $h_2$ at the LHC, we need to quantify its total decay width and its branching ratios.
Scalar mixing allows the $h_2$ to decay into SM particles with the same partial decay widths as a would-be SM Higgs boson of the same mass, multiplied with $s_\alpha^2$.
If $m_{h_2} > 2\, m_{h_1} \simeq 250$ GeV, the $h_2$ can also decay into the SM-like Higgs boson $h_1$. The corresponding partial decay width is given by
\begin{equation}
\Gamma_{h_1 h_1} =  \frac{s_\alpha c^2_\alpha (2 m_{h_1}^2 + m^2_{h_2})}{32 \pi m_{h_2} v_{\rm EW}}\sqrt{1-\frac{4\, m_{h_1}^2}{m_{h_2}^2}}\,.
\end{equation}
Furthermore, the coupling to $N_1$ and $N_2$ allows for decays into heavy and light neutrinos. The decay width into two heavy neutrinos is
\begin{equation}
\Gamma_{NN} = \frac{c_\alpha^2\,m_{h_2}\, y_H^2}{4\pi} \left(1 - \frac{4\, M^2}{m_{h_2}^2}\right)^{3\over 2} \,,
\end{equation}
while the decays of the $h_2$ that involve light neutrinos are suppressed by the small active-sterile neutrino mixing parameter $|\theta|^2$ and thus neglected.

We show in the left panel of fig.\ \ref{fig:h2-rvis} the resulting total decay width for the $h_2$, for three representative values of $y_H$, assuming that the heavy neutrino masses are given by $y_H\,v_H$.
The right panel of fig.\ \ref{fig:h2-rvis} shows the branching ratio of the $h_2$ into SM particles, which are dominantly $W,\,Z$, the top quark and the $h_1$ boson, with the approximate individual branching ratios being 55, 30, 15 and $\sim 1$\%, respectively, for $y_H \ll 0.1$ and $m_{h_2}> 350$ GeV. When the coupling $y_H\sim 0.1$, the branching ratio of $h_2$ into heavy neutrinos is $\sim$50\%, and its branching into SM particles is approximately halved, whereas for $y_H = {\cal O}(1)$, the $h_2$ dominantly decays into heavy neutrinos, and its branching into SM particles is suppressed by a factor ${\cal O}(100)$.
It is interesting to note that $y_H = 1$ can lead to a total decay width as large as $\Gamma_{\rm tot} = {\cal O}(100)$ GeV for $m_{h_2} \sim 1$ TeV. In this case, the $h_2$ decays into heavy neutrinos with a branching ratio between 96\% to 99.7\%.

We remark that the heavy neutrinos themselves might constitute an invisible decay channel, but not necessarily so: due to their admixture of left-handed neutrinos, they may decay inside the detector via the weak currents and the Higgs boson. An interesting signature at the LHC would be a di-heavy-neutrino, which could be a double-semileptonic final state, $(\ell^\pm_\alpha jj)\,(\ell^\pm_\beta j j)$ or double-leptonic final state, $(\ell^\pm_\alpha \ell^\mp_\alpha)\,(\ell^\pm_\beta \ell^\mp_\beta)$ (each with missing transverse energy), with a clear angular separation.

We notice, that it may be possible to infer the corresponding heavy neutrino mass scale via the branching ratios of the $h_2$ and its mass.
Should the heavy scalar boson be found at the LHC, e.g.\ via a significant excess in $W$ and $Z$ bosons (and top quarks) with invariant mass $m_{h_2}$, the scalar mixing is fixed by the condition \eqref{eq:condition} and the total number of $h_2$ events for luminosity ${\cal L}$ is predicted: $N_{h_2}^{\rm tot} = \sigma_{h_2}(m_{h_2})\, {\cal L}$. 
This allows a determination of the $h_2$ branching ratio into two $W$ bosons,
\begin{equation}
{\rm Br}(h_2 \to WW) = \frac{N_{h_2}^{W}}{N_{h_2}^{\rm tot}}\,,
\end{equation}
with $N_{h_2}^{W}$ being the number of observed $W$ events. 
When the masses (and thus the scalar mixing) are known, the partial decay widths contributing to $\Gamma_{\rm tot}$ are fixed, and it is possible to extract the value of $\Gamma_{NN}$ from $\Gamma_{\rm tot}$, which in turn fixes the masses of the heavy neutrinos through the parameter $y_H$.

\section{Two case studies}
In this section we will conduct a case study for two explicit values of the heavy scalar mass $m_{h_2}$, which are motivated from LHC results. In this line we interpret each, the recent diphoton excess \cite{diphoton} and the statistical signal with three sigma significance reported in ref.\ \cite{vonBuddenbrock:2015ema} as the first hints for a heavy scalar with a mass of 272 and 750 GeV, respectively, which we refer to as $m_{272}$ and $m_{750}$ in the following.

The corresponding production cross section for the past and present run of the LHC, as shown in fig.\ \ref{fig:h2production}, are given by:
\begin{subequations}
\begin{eqnarray}
\sigma_{m_{272}} & = & \left\{ \begin{array}{cc} 9.6 \pm 7.0 \text{ fb}  & \text{8 TeV}\,\, \\ 29 \pm 21 \text{ fb} & \text{13 TeV}\,, \end{array}\right.\\
\sigma_{m_{750}} & = & \left\{ \begin{array}{cc}  68 \pm 49  \text{ ab} & \text{8 TeV}\,\, \\ 0.31 \pm 0.22 \text{ fb} & \text{13 TeV}\,. \end{array}\right.
\end{eqnarray}
\end{subequations}
The magnitude of the cross section $\sigma_{m_{272}}$ suggests that ${\cal O}(10^3)$ events are to be expected in the 8 TeV data. On the other hand, by the end of data taking around ${\cal O}(10^5)$ $h_2$ bosons may have been produced in  3 ab$^{-1}$ at $\sqrt{s}=13$ TeV, and even up to ${\cal O}(10^7)$ events would be possible at the FCC-hh, which may be sufficient for a detection.

Concerning the $m_{750}$ we notice that the cross section $\sigma_{m_{750}}$ leads to less than one expected event at the 8 TeV LHC, and ${\cal O}(1)$ event in the 13 TeV run, which is about an order of magnitude too small to account for the observed diphoton excess, ignoring the fact that Br$(h_2\to \gamma\gamma)\leq  2 \times 10^{-7}$ for $m_{h_2}=750$ and $y_H \ll 0.1$.
Moreover, the large decay width, $\Gamma_{750}\sim 50$ GeV, requires $y_H\sim 1$, in which case the $h_2$ decays into heavy neutrinos with a branching ratio of $\geq 96\%$, such that it is impossible to explain the diphoton excess in this framework.
By the end of data taking, $m_{750}$ might yield up to ${\cal O}(10^3)$ events for 3 ab$^{-1}$ at $\sqrt{s}=13$ TeV, and at the FCC-hh one would expect up to ${\cal O}(10^6)$ events.
Whether or not the LHC will have prospects for the discovery of a signal in $W$ and $Z$ bosons, or, possibly, of the exotic signature of a di-heavy-neutrino, requires a dedicated analysis, which is beyond the scope of this article.

An upper bound for the heavy neutrino mass scale can be inferred from fig.\ \ref{fig:Hillvev} via the vacuum expectation value, which is $v_H \leq 89.2$ GeV and $v_H \leq 246.0$ GeV for $m_{272}$ and $m_{750}$, respectively. This implies that for $m_{272}$ and $m_{750}$ the resulting Majorana mass scale makes the heavy neutrinos kinematically available for production at any of the planned lepton colliders, the ILC \cite{Baer:2013cma}, CLIC \cite{Battaglia:2004mw}, FCC-ee \cite{Gomez-Ceballos:2013zzn}, or CEPC \cite{preCDR}, and also at the electron-proton collider LHeC \cite{AbelleiraFernandez:2012cc}, which thus have very promising prospects of discovering the heavy neutrinos.

\section{Conclusions}
The Standard Model is extraordinarily successful in describing precision data at all accessible energy scales, but it fails by design to generate the light neutrinos' masses. Moreover, the metastability of the electroweak vacuum suggests, that there is a higher energy scale that is connected to new physics.

In this article we studied the extension of the SM with the scalar Hill field and two right-handed neutrinos, which yields a minimal framework to address these shortcomings of the SM, without spoiling the agreement between theory and precision measurements.
We imposed the condition that the SM-like Higgs boson's self coupling and its running vanish at the Planck scale, which fixes the scalar mixing via the mass of a second scalar ($h_2$), that is heavier than the SM-like Higgs boson. This condition removes the necessity of additional scales in the theory, such that we consider the $h_2$ mass to be on the weak scale, i.e.\ $m_{h_2}\leq 1$ TeV.
In this framework, the Majorana masses are generated dynamically when the Hill field develops its vacuum expectation value, which is proportional to $m_{h_2}$.

We have assessed the production and decay properties of the $h_2$ at the LHC, and we pointed out the intriguing possibility to infer the heavy neutrinos' mass scale from its total width, its branching ratios into SM particles, and into heavy neutrinos, which may or may not be detectable.

As a case study we have fixed $m_{h_2}$ according to hints in recent LHC analyses: a statistical signal at 272 GeV, and the diphoton excess at 750 GeV.
If $h_2$ has a mass of 272 GeV it may be possible to confirm a discovery at the LHC with 3 ab$^{-1}$, provided that it does not decay mainly into invisible (i.e.\ very weakly mixed) heavy neutrinos.
Interestingly, $h_2$ with a mass of $750$ GeV accommodates a width comparable to that of the diphoton excess ($\sim 50$ GeV) when its couplings to the heavy neutrinos are ${\cal O}(1)$. Since this implies the $h_2$ branching ratio into heavy neutrinos to be at least 96\%, it is, however, not possible to explain the diphoton excess in this framework.

We conclude that the observation of a scalar resonance at the LHC or future hadron colliders and its decay properties might yield clues for the mass scale of the neutrino sector, and vice versa.

\subsection*{Acknowledgements}
This work has been supported by the Swiss National Science Foundation. I want to thank Stefan Antusch for his careful reading of the manuscript and invaluable feedback, and also Hyung Do Kim for useful discussions.

\bibliographystyle{unsrt}

\end{document}